\documentclass[%
 reprint,
showpacs,
 amsmath,amssymb,
 aps,
floatfix,
]{revtex4-1}

\usepackage{graphicx}% Include figure files
\usepackage{braket}
\usepackage{float}
\DeclareMathOperator{\sinc}{sinc}

\begin{document}

\title{Spontaneous Parametric Down-Conversion in asymmetric couplers: \\
photon purity enhancement and intrinsic spectral filtering}% Force line breaks with \\

\author{Philip B. Main}
\affiliation{
 Centre for Photonics and Photonic Materials, 
 Department of Physics, University of Bath, Bath BA27AY, UK}%Lines break automatically or can be forced with \\

\author{Peter J. Mosley}
 \affiliation{
 Centre for Photonics and Photonic Materials, 
 Department of Physics, University of Bath, Bath BA27AY, UK}

\author{Andrey V. Gorbach}
 \email{A.Gorbach@bath.ac.uk}
 \affiliation{
 Centre for Photonics and Photonic Materials, 
 Department of Physics, University of Bath, Bath BA27AY, UK}

\date{\today}% It is always \today, today,

\begin{abstract}
We analyze the process of photon-pair generation via spontaneous parametric down-conversion in a quadratic nonlinear asymmetric waveguide coupler. The two waveguides have different geometry, such that light coupling only occurs within a narrow bandwidth of one of the generated (signal) photon modes, while the other (idler) photon together with the pump stay localized in one (driven) arm of the coupler. We demonstrate that such a setup represents a powerful and flexible tool for engineering spectral properties of generated photon pairs. Mode hybridization and dispersion of coupling can be utilized for shifting the balance between group velocities of interacting pump, signal and idler fields, subsequently leading to a significant increase of spectral factorisability (purity) of photons. We also show that for interaction lengths shorter than one beat length, generated pairs with signal photon being localized in the auxiliary (not pumped) arm of the coupler appear to be spectrally localized in both signal and idler components. The bandwidth of such intrinsic filtering of generated photons can be controlled by several geometrical parameters.
\end{abstract}

\maketitle

%\tableofcontents

\section{Introduction}

Photons are lauded as one of the few quantum systems able to maintain their coherence at room temperature, making them an ideal candidate for quantum technologies \cite{Knill2001, Kok2007a, OBrien2009}.
A considerable progress in establishing such schemes on integrated optics platforms have been achieved in the past decade \cite{Politi2008, Giovannetti2011, Silverstone2014a, Meany2015, Luo2019}.

Optical waveguide couplers and arrays are important elements of classical and quantum optical circuits. Photon "hopping" between adjacent waveguides represents an important degree of freedom, that has been exploited e.g. for multi-photon path entanglement \cite{Matthews2009a}, development of quantum logic gates \cite{Politi2008, Meany2015, Crespi2011}, and quantum walks \cite{Peruzzo2010c}. Furthermore, by utilising intrinsic quadratic ($\chi_2$) or Kerr ($\chi_3$) nonlinearities of the waveguide, one can integrate heralded single photon sources based on Spontaneous Parametric Down-Conversion (SPDC) \cite{Tanzilli2001b, Banaszek2001} or Spontaneous Four-Wave Mixing (SFWM) \cite{Li2004, Harada2011, Silverstone2014a}, respectively, into a circuit. Development of such on-chip integrated single-photon sources has attracted much attention recently. In particular, integration of sources with waveguide arrays \cite{Solntsev2012a} has led to the establishment of the concept of driven quantum walks \cite{Hamilton2014a} and a promising platform for on-chip generation of path-entangled photon states \cite{Solntsev2017}. All-‐optical control of the output quantum state and photon pair steering has been recently demonstrated in a quadratic nonlinear waveguide coupler \cite{Setzpfandt2016}.

\begin{figure}
    \includegraphics[width=0.49\textwidth]{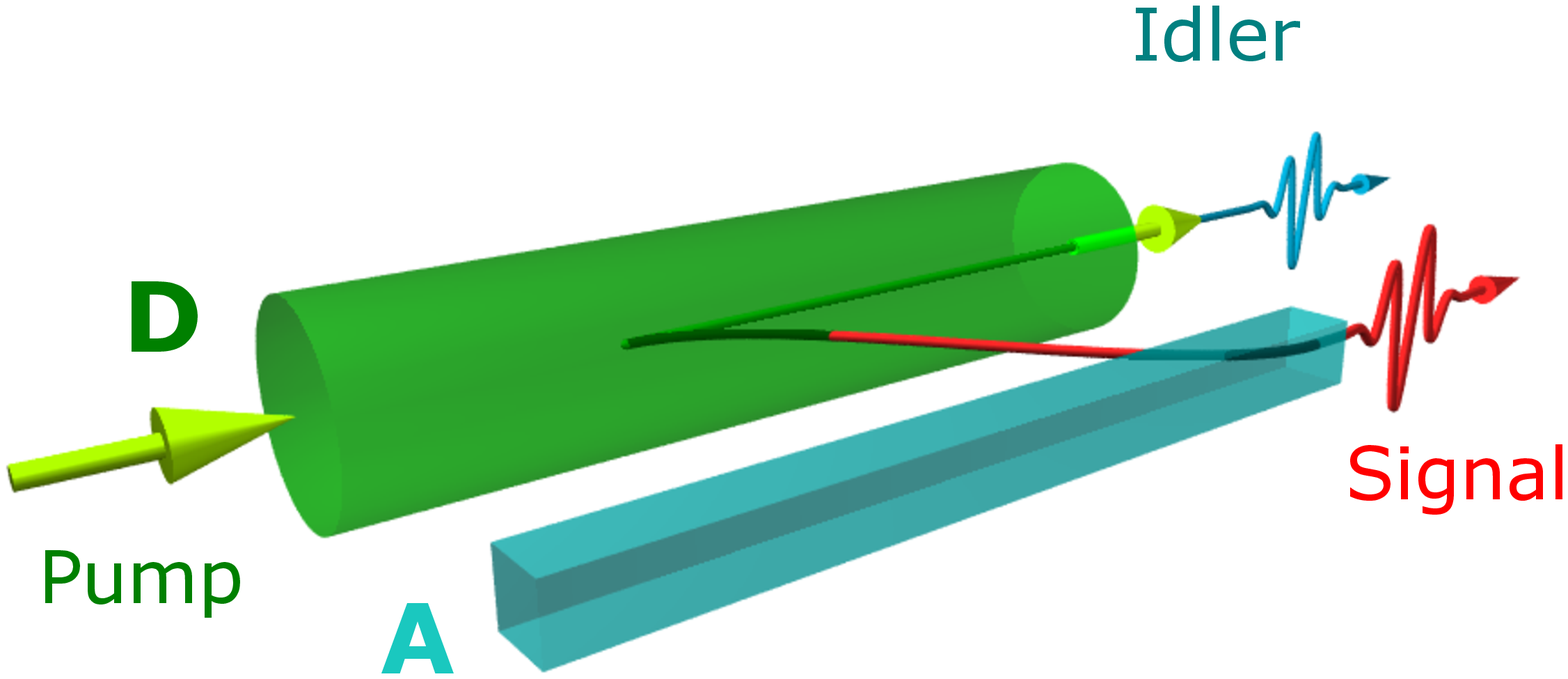}
    \caption{(Color online) A scheme of SPDC in an assymetric coupler: pump and idler remain confined in the driven ($D$) waveguide, signal photon can couple to the auxiliary ($A$) waveguide.}
   \label{fig:scheme}    
\end{figure}

Most studies hitherto have been focused on photon pair generation in couplers and arrays of identical waveguides.
In such systems generated signal and idler photons are simultaneously free to walk across the array.
In this work, we consider the SPDC-driven photon-pair generation process in an asymmetric waveguide coupler with two adjacent waveguides of different geometries. While two dissimilar waveguides will generally have different sets of modes, photon "hopping" will be effectively suppressed except at specific wavelengths where matching of propagation constants of a pair of modes is engineered. Particularly, we consider a setup where a bright light pump and one of the generated photons (idler) remain confined within the driven waveguide (D), while the other photon (signal) can couple to the adjacent auxiliary waveguide (A), see Fig.~\ref{fig:scheme}. 

We discuss two different applications of such a setup as a source of heralded single photons: whereby the heralding is done by detecting either idler photon in waveguide D, or signal photon in waveguide A. In the former case, mode hybridization and dispersion of coupling can be utilized for local adjustments of the signal photon dispersion, without affecting properties of the pump and idler fields. Such flexibility is crucial for controlling spectral factorisability of generated photon pairs. In the latter case, spectrally local coupling combined with a specific interference of the two signal modes can lead to a pronounced modification of the joint spectral properties of generated photon pairs and intrinsic spectral filtering.

\section{Two-photon state function}

\begin{figure*}
\includegraphics[width=0.9\textwidth]{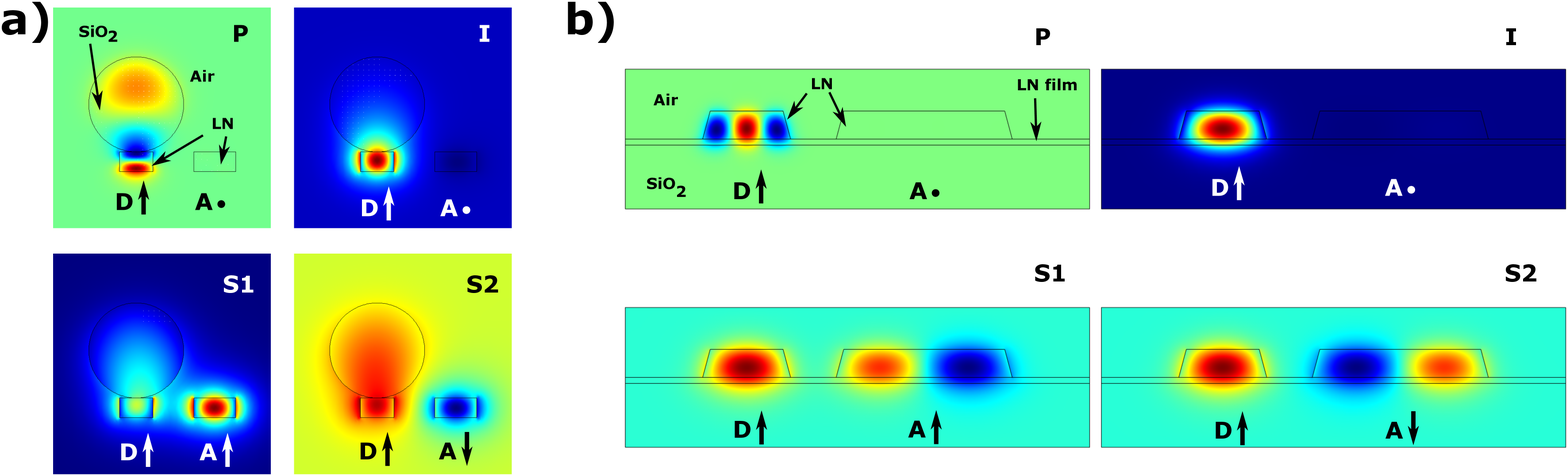}
\caption{(Color online) Example asymmetric coupler structures. $P$, $I$, $S1$ and $S2$ mode profiles (dominant electric field component) in: (a) a hybrid microfibre-lithium niobate (LN) waveguide \cite{Gorbach2015a, Main2016, Cai2018} (a microfibre of diameter $D=1.44\mu$m ontop of a $510 nm \times 300 nm$ LN waveguide, waveguide $D$ ) adjacent to a suspended $635 nm \times 300 nm$ LN waveguide (waveguide $A$), edge-to-edge separation $620$nm, $\lambda_p=0.72\mu$m, $\lambda_i=1.34\mu$m, $\lambda_s=1.55\mu$m; (b) $1200nm\times460nm$ (waveguide $D$) and $2650nm\times460nm$ (waveguide $A$) LN ridge waveguides \cite{Luo2019, Desiatov2019a}, edge-to-edge separation $750$nm, LN film thickness $100$nm, $\lambda_p=775$nm, $\lambda_i=1400$nm, $\lambda_s=1735$nm. In both examples S1 and S2 supermodes can be considered as in-phase and anti-phase super-positions of modes of individual waveguides D and A.}
\label{fig:modes}
\end{figure*}

In SPDC process a photon from bright pump ($\omega_p$) is spontaneously converted into a pair of signal and idler photons, such that $\omega_p=\omega_s+\omega_i$. Hereafter we assume that the pump and idler fields are excited in waveguide D modes only (modes $p$ and $i$, respectively), while the signal photon can be excited in two hybridized modes of the coupler, $s1$ and $s2$, see Fig.~\ref{fig:modes}. This process can thus be described by the following interaction Hamiltonian \cite{Mandel1995}:

\begin{eqnarray}
\nonumber
 H_I(z) &=&  \iint d\omega_i d\omega_s \: \left\{A_p(\omega_s + \omega_i,z)\hat{a}_{i,H}^\dagger(\omega_i,z)\times\right. \\ 
 &&\left.\left[  \gamma_{s1} \hat{a}_{s1,H}^\dagger(\omega_s,z) 
 + \:  \gamma_{s2} \hat{a}_{s2,H}^\dagger(\omega_s,z) \right] \right\}+ h.c.\:
\label{eq:Hint}
\end{eqnarray}
where $A_p(\omega,z)=\alpha(\omega)e^{i\beta_p(\omega) z}$ is pump field, its spectral content is described by the pump function $\alpha(\omega)$, 
$\beta_{p,s1,s2,i}$ are propagation constant of the modes,
$\hat{a}_j$ ($\hat{a}_j^\dagger$) are photon annihilation (creation) operators in the $j$th mode of the structure, $j=s1,s2,i$, signal and idler modes propagation is encapsulated in $z$-dependence of the operators 
$\hat{a}_{j,H}(\omega,z)=\hat{a}_{j}(\omega)e^{i\beta_{j}z}$, $\hat{a}_{j,H}^\dagger(\omega,z)=\hat{a}_{j}^\dagger(\omega)e^{-i\beta_{j}z}$, coefficients $\gamma_{s1,s2}$ are determined by the overlap integrals of the corresponding interacting modes $(p,s1,i)$ and $(p,s2,i)$, respectively \cite{Main2016}.

Setting vacuum state as the initial condition, and assuming sufficiently weak nonlinearity, the state vector can be approximated as \cite{Rubin1996, Grice1997a, Couteau2018}:
\begin{eqnarray}
&&\ket{\psi(z)}  
 = e^{-i\int_0^z \hat{H}_I(z^{\prime})dz^\prime} \ket{\textrm{vac}}    
\nonumber 
\\
&&\:\approx 
 \left[ 1  - i\int_0^z \hat{H}_I(z^\prime) dz^\prime \right]\ket{\textrm{vac}} = \ket{\textrm{vac}} + \ket{\psi_2}\;.
\label{eq:psi2_expand}
\end{eqnarray}
In the above expansion, the two-photon state function $\ket{\psi_2}$ emerges as the first non-trivial term:
\begin{equation}
\begin{split}
\ket{\psi_2(z)} =  \iint d\omega_s d\omega_i \: \{\alpha(\omega_s + \omega_i)\ket{\omega_i} \times \\
\times  \sum_{m} \Phi_{m}(\omega_s, \omega_i, z) \ket{\omega_s, m}
\}.
\end{split}
\label{eq:psi2}
\end{equation}
Here the summation is performed over the signal hybridized modes $m=(s1,s2)$, $\ket{\omega_s,(s1,s2)}=\hat{a}^\dagger_{(s1,s2)}(\omega_s)\ket{\textrm{vac}}$, $\ket{\omega_i}=\hat{a}^\dagger_{i}(\omega_i)\ket{\textrm{vac}}$ are single photon states in the signal and idler modes at the respective frequencies, and the phase-matching functions are:
\begin{eqnarray}
\nonumber
\Phi_{m}(\omega_s, \omega_i, z) & = & -i \int_0^z  \: \gamma_{m} \: e^{i \Delta \beta_{m} z^\prime} dz^\prime\\
&=& -i z \:\gamma_{m} \: e^{\frac{i \Delta \beta_{m} z }{2}} \textrm{sinc}\left(\frac{\Delta \beta_m z }{2}\right) \;,
\label{eq:Phi_j}
\\
\Delta \beta_{m}(\omega_s,\omega_i) & = & \beta_p(\omega_s+\omega_i) - \beta_{m}(\omega_s) - \beta_i(\omega_i)\;.
\end{eqnarray}

\begin{figure}
\includegraphics[width=0.48\textwidth]{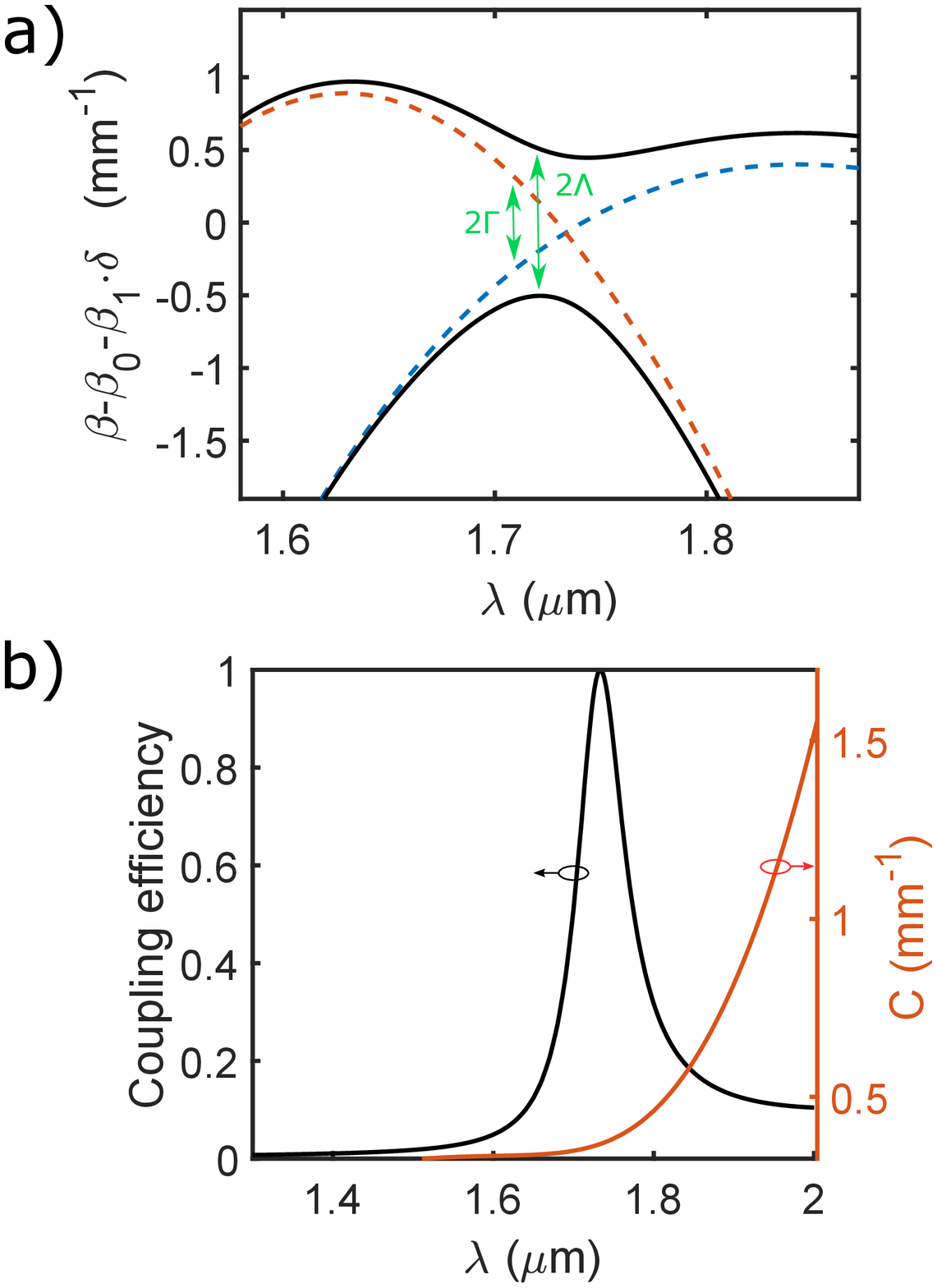}
\caption{ (Color online) Mode hybridization in the example structure as in Fig.~\ref{fig:modes}(b): (a) propagation constants of isolated waveguide D (dashed blue) and waveguide A (dashed red) anti-cross in a vicinity of $\lambda_s=1735$nm. The resulting S1 and S2 super-modes are shown with solid curves. For clarity of presentation, shifted propagation constants are plotted $\widetilde{\beta}=\beta-\beta_0-\beta_1\delta$, where $\beta_1=(\beta^\prime_D+\beta^\prime_A)/2$ is the average inverse group velocity of isolated waveguides D and A at the wavelength of anti-crossing; (b) the coupling efficiency $\eta_c$ and coupling coefficient $C$ calculated from the splitting of the two modes ($\Lambda$ and $\Gamma$ parameters).\label{fig:mode_dispersion}}
\end{figure}

It is instructive to consider the two-photon state in the basis of modes of individual waveguides D and A. Mode hybridization of the coupler can be effectively described by a generic two-level system model (see Appendix A for more detail). The change of basis can be represented by a unitary transformation:

\begin{eqnarray}
\begin{bmatrix}
\hat{a}_D(\omega,z) \\
\hat{a}_{A}(\omega,z)
\end{bmatrix}
=
e^{i\beta_0 z} 
\hat{T}
\begin{bmatrix}
e^{i\Lambda z} \hat{a}_{s1}(\omega, 0) \\
e^{-i\Lambda z} \hat{a}_{s2}(\omega, 0)
\end{bmatrix}
\;, 
\label{eq:basis_transform}
\\
\hat{T} = \frac{1}{\sqrt{2\Lambda}}
\begin{bmatrix}
\sqrt{\Lambda + \Gamma} & \sqrt{\Lambda - \Gamma} \\
\sqrt{\Lambda - \Gamma} & -\sqrt{\Lambda + \Gamma} \\
\end{bmatrix}\;,
\label{eq:T_matrix}
\end{eqnarray}
where $\beta_0=(\beta_{s1}+\beta_{s2})/2$, $\Lambda=(\beta_{s1}-\beta_{s2})/2$, $\Gamma=(\beta_{D}-\beta_{A})/2$, and $\beta_{D,A}$ are propagation constants of the signal modes in isolated D and A waveguides, see Fig.~\ref{fig:mode_dispersion}(a).

In the basis of modes of individual waveguides D and A, the two-photon state will have the same functional form as in Eq.~(\ref{eq:psi2}), where the summation is now performed over the modes $m=A,D$, and the corresponding phase-matching functions are given by:

\begin{eqnarray}
\Phi_{D}
&=&  \left[\sqrt{\frac{\Lambda + \Gamma}{2\Lambda}} e^{i\Lambda z} \Phi_{s1} + \sqrt{\frac{\Lambda - \Gamma}{2\Lambda}}e^{-i\Lambda z}\Phi_{s2}\right]\;,
\label{eq:PhiD}
\\
\Phi_{A} &=&  \left[\sqrt{\frac{\Lambda - \Gamma}{2\Lambda}} e^{i\Lambda z} \Phi_{s1} - \sqrt{\frac{\Lambda + \Gamma}{2\Lambda}}e^{-i\Lambda z}\Phi_{s2}\right] \;.  
\label{eq:PhiA}
\end{eqnarray}

We note that the coupling coefficient in the two-level model $C=\sqrt{\Lambda^2-\Gamma^2}$ generally increases with wavelength, due to the increasing overlap of evanescent fields of the modes of individual waveguides. However this only affects the beat length between the two super-modes $L_c=\pi/C$, while the fraction of power being transferred between the two waveguides over one period (coupling efficiency):
\begin{equation}
 \eta_c=1-\frac{\Gamma^2}{\Lambda^2}\;,
 \label{eq:coupling_eff}
\end{equation}
(see Appendix A for details). In an asymmetric coupler, the coupling efficiency reaches its maximum value of $1$ at the wavelength of anti-crossing (where $\Gamma=0$), and decays away from this point (as $\Gamma/\Lambda \to \pm 1$, note also that $\Gamma$ has opposite signs on either side of the anti-crossing point), see Fig.~\ref{fig:mode_dispersion}. As the result, phase matching functions $\Phi_D$ and $\Phi_A$ asymptotically converge to either $\Phi_{s1}$ or $\Phi_{s2}$ (with a constant pre-factor $(1/\sqrt{2})e^{\pm i\Lambda z}$) away from the hybridization region. However, in a vicinity of the anti-crossing, interference between $\Phi_{s1}$ and $\Phi_{s2}$ terms in Eqs.~(\ref{eq:PhiD}), (\ref{eq:PhiA}) creates the important difference in properties of two-photon states when measured in $(s1,s2)$ and $(D,A)$ bases.

\section{spectral purity enhancement by mode hybridization}

Photon pairs generated by SPDC will generally emerge with a pronounced spectral anti-correlation. This entanglement, however is not always a favourable property of photon-pair states. In particular, for the development of SPDC-based single photon sources, whereby generation of a signal photon is heralded by detection of its idler "partner", factorable states are most useful because the heralded photon remains in a pure quantum state \cite{Meyer-Scott2017}. As follows from Eq.~(\ref{eq:psi2}), the correlated quantum state in a particular combination of signal and idler modes is characterized by the joint spectral amplitude (JSA):
\begin{equation}
f_m(\omega_s,\omega_i)=\alpha(\omega_s+\omega_i)\Phi_m(\omega_s,\omega_i)\;,
    \label{eq:JSA}
\end{equation}
which is the product of pump and phase matching functions. Factorisability of the signal-idler states
can be characterised by the spectral purity $P=\textrm{Tr}(\rho_i^2)$, where $\rho_i=\textrm{Tr}_i(\ket{\psi_2}\bra{\psi_2})$ is the reduced density matrix obtained by partial trace over the idler degrees of freedom \cite{Cassemiro2010}. It can be calculated from the expansion coefficients of JSA function into the weighted sum of separable functions (Schmidt decomposition) $f(\omega_s,\omega_i=\sum_k \sqrt{b_k} u_k(\omega_s)v_k(\omega_i)$: $P=\sum b_k^2/(\sum b_k)^2$, $0\le P \le 1$ with $P=1$ corresponding to a pure heralded photon state \cite{Law2000,Braunstein2005}. The intrinsic anti-correlation dictated by the energy conservation principle $\omega_i=\omega_p-\omega_s$, embedded in the pump spectral function $\alpha(\omega_s+\omega_i)$, can be compensated by a properly tuned phase-matching function $\Phi$. Expanding $\Phi$ in a vicinity of a phase-matching point $\Delta\beta(\omega_{s0},\omega_{i0})=0$ leads to:
\begin{equation}
    \Phi\sim \textrm{sinc}\left[\frac{(\beta_p^\prime-\beta_s^\prime)\delta_s z +(\beta_p^\prime-\beta_i^\prime)\delta_i z}{2}\right]\;,
\end{equation}
where $\delta_{s,i}=\omega_{s,i}-\omega_{s0,i0}$, $\beta_j^\prime=\partial \beta_j/\partial \omega_j (\omega_{j0})$ are inverse group velocities of the pump, signal and idler at the phase matching point $\omega_{p0}=\omega_{s0}+\omega_{i0}$. A positive correlation in $\Phi$ can be achieved when:
\begin{equation}
\frac{1-(\beta_s^\prime/\beta_p^\prime)}{1-(\beta_i^\prime/\beta_p^\prime)}<0\;,
\label{eq:good_purity}
\end{equation}
in other words when the inverse group velocities are related to each other as $\beta_i^\prime<\beta_p^\prime<\beta_s^\prime$ \cite{Edamatsu2011}. However, this is not often possible to achieve in simple waveguide geometries with only few geometrical parameters (such as waveguide width and height) available for dispersion engineering.

\begin{figure}
    \includegraphics[width=0.4\textwidth]{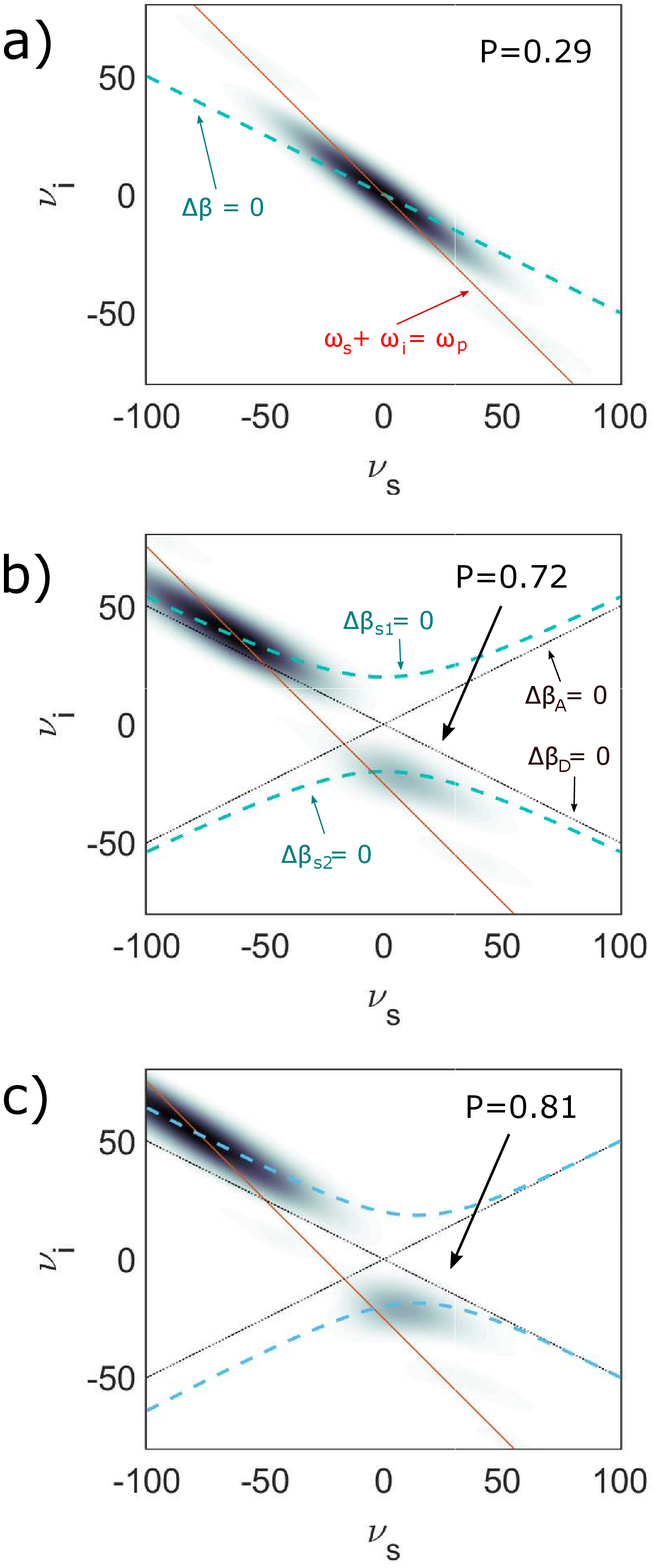}
    \caption{(Color online) JSA (absolute value) as function of dimensionless frequency detunings $\nu=\delta\cdot(\beta_p^\prime L/\pi)$ at a fixed propagation distance $L$ in an isolated waveguide D (a), waveguide coupler with $C\cdot L/\pi =2$, $C^\prime =0$ (b) and $C^\prime / \beta_p^\prime=-0.02$ (c). The corresponding calculated purities are stated in the top right corner of each picture. Disperson parameters of the driven D and auxiliary A waveguides at the phase-matching point $\omega_{s0}$ are: $\beta_{sD}=\beta_{sA}$, $\beta_{sD}^\prime/\beta_p^\prime=1.05$, $\beta_{sA}^\prime/\beta_p^\prime=0.95$, $\beta_i^\prime/\beta_p^\prime=1.1$. Energy and momentum conservation conditions are indicated with solid and dashed lines, respectively.}
    \label{fig:purities}
\end{figure}

The addition of an auxiliary waveguide (A) represents a powerful yet relatively simple method of local manipulation of the signal mode dispersion, while keeping pump and idler practically unaffected. Mode hybridization between the two waveguides induces strong variation of $\beta_s^\prime$, controllable by the auxiliary waveguide dispersion in the vicinity of signal frequency range $\beta_A(\delta_s)\approx \beta_{sA}+\beta_{sA}^\prime \delta_s$, the coupling coefficient $C=\sqrt{\Lambda^2-\Gamma^2}$ (tuned by the waveguide separation), and its dispersion [see Eq. (A3) in the Appendix]:

\begin{eqnarray}
\nonumber
 \beta_{s1,s2}^\prime&=&\frac{\beta_{sD}^\prime+\beta_{sA}^\prime}{2}\\ 
 &&\pm \frac{\left(\beta_{sD}-\beta_{sA}\right)\left(\beta_{sD}^\prime-\beta_{sA}^\prime\right)+4CC^\prime}{2\sqrt{\left(\beta_{sD}-\beta_{sA}\right)^2+4C^2}}
 \end{eqnarray}

Assuming that the modes in waveguides $D$ and $A$ are matched at $\omega_{s0}$, $\beta_{sD}=\beta_{sA}$, the two main control parameters are auxiliary mode dispersion $\beta_{sA}^\prime$ and coupling dispersion $C^\prime$. In Fig.~\ref{fig:purities} we illustrate how each of the two parameters can be utilised to improve photon purity. In our calculations we used a Gaussian pump function:
\begin{equation}
\alpha(\omega)=\alpha_0 \exp\left[-\frac{(\omega-\omega_p)^2}{2\sigma^2}\right]\;,
\label{eq:pump_function}
\end{equation}
where $\alpha_0$ is a scaling factor proportional to the pump peak power (it does not affect calculations of purity), pump central frequency $\omega_p$ and spectral width $\sigma$ were adjusted in each geometry to maximize purity factor $P$.

For waveguide D we select $\beta_{sD}^\prime=1.05\beta_p^\prime$ and $\beta_i^\prime=1.1\beta_p^\prime$, such that the condition in Eq.~(\ref{eq:good_purity}) is not met. The corresponding JSA function in Eq.~(\ref{eq:JSA}) is shown in Fig.~\ref{fig:purities}(a). It displays a strong anti-correlation, and the resulting purity is low: $P\approx 0.29$. Adding an auxiliary waveguide with $\beta_{sA}^\prime=0.95\beta_p^\prime$, it is possible to boost the purity by more than a factor of two for photon pairs generated in the $S2$ super-mode in the vicinity of the anti-crossing, see Fig.~\ref{fig:purities}(b). We emphasise that the efficiency of photon-pair generation is determined by the amplitude of a particular signal mode in the driven $D$ waveguide, where the pump is localized. Neglecting weak interactions due to evanescent pump field in A waveguide, nonlinear coefficients $\gamma_{1,2}$ which determine efficiency of photon pair generation in S1 and S2 signal supermodes, respectively,  can be approximated as:
\begin{equation}
\gamma_{s1}=\sqrt{\frac{\Lambda+\Gamma}{2\Lambda}}\gamma_0\;, \qquad \gamma_{s2}=\sqrt{\frac{\Lambda-\Gamma}{2\Lambda}}\gamma_0\;,
\label{eq:gam1and2}
\end{equation}
where $\gamma_0$ is the nonlinear coefficient in isolated D waveguide.
Thus compared to isolated $D$ waveguide, the efficiency is reduced by a factor of $\sim \sqrt{2}$ in $S1$ and $S2$ super-mode in a vicinity of the anti-crossing (where $\Gamma$ tends to zero). By tuning the pump central frequency $\omega_p$, it is possible to arrange phase matching further away from the anti-crossing, and increase the resulting purity even further (i.e. where $S1$ or $S2$ super-modes converge to the mode of isolated waveguide $A$), however at the cost of a significantly lower efficiency (as $\Gamma$ tends to $\pm\Lambda$, see Fig.~\ref{fig:mode_dispersion}).

Dispersion of coupling can also assist in boosting the purity, as illustrated in Fig.~\ref{fig:purities}(c). Generally, coupling is stronger at larger wavelength due to stronger overlap of evanescent fields, leading to $C^\prime<0$. Therefore it decreases (increases) $\beta_s^\prime$ for $S1$ ($S2$) super-mode, and shifts the balance in Eq.~(\ref{eq:good_purity}).

\section{Intrinsic spectral filtering of photon pairs}
 
In previous section it was assumed that heralding is performed using idler photon, which is confined in the driven D waveguide. Here we focus on a different setup where the signal photon is used for heralding, and the photon detector is connected to the auxiliary A waveguide. Noteably the two cases are not symmetrical, since the signal photon is generated in the $S1$ and $S2$ super-modes of the structure. Therefore, unlike the idler photon, it is not confined to a particular waveguide. Performing detection in the waveguide A only, the signal photon state is collapsed to a particular combination of the S1 and S2 super-modes. The process is best described in the basis of individual waveguide modes. 

\begin{figure}
\includegraphics[width = 0.44\textwidth]{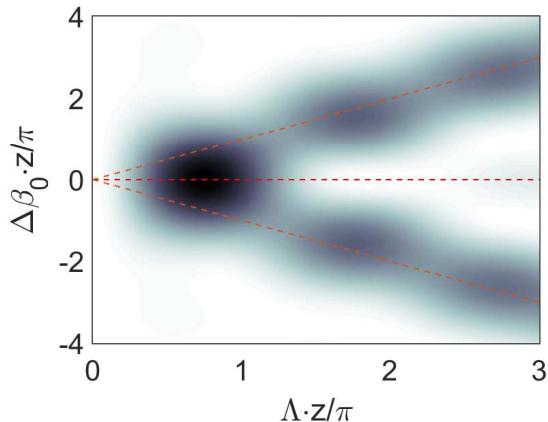}
\caption{(Color online) Phase function $\Phi_A$ (absolute value) as a function of $\Lambda z$ and $\Delta \beta_0 z$ with fixed values of $\Lambda$ and $\Gamma$. Dashed lines indicate asymptotes $\Delta\beta_0=0,\pm\Lambda$. For clarity of illustration, the overall linear growth of $\Phi_A$ with $z$ is removed.}
\label{fig:filter}
\end{figure}

Adopting the expressions for nonlinear coefficients in Eq.~(\ref{eq:gam1and2}), the phase-matching functions in Eqs.~(\ref{eq:PhiD}) and (\ref{eq:PhiA}) can be written as:
\begin{widetext}
\begin{eqnarray}
\Phi_{D} = -iz\gamma_0\frac{1}{2\Lambda} e^{i\frac{\Delta \beta_0 z}{2}}\left[(\Lambda + \Gamma) e^{i\frac{\Lambda z}{2}}\sinc \left( \frac{(\Delta \beta_0 - \Lambda) z}{2} \right) + (\Lambda - \Gamma)e^{-i\frac{\Lambda z}{2}}\sinc \left( \frac{(\Delta \beta_0 + \Lambda) z}{2} \right)\right] \;,\\
\Phi_{A} = -iz\gamma_0\frac{\sqrt{\eta_c}}{2} e^{i\frac{\Delta \beta_0 z}{2}}\left[ e^{i\frac{\Lambda z}{2}}\sinc \left( \frac{(\Delta \beta_0 - \Lambda) z}{2} \right) - e^{-i\frac{\Lambda z}{2}}\sinc \left( \frac{(\Delta \beta_0 + \Lambda) z}{2} \right)\right] \;,
\end{eqnarray}
\end{widetext}
where $\Delta\beta_0=\beta_p-\beta_i-\beta_0$ with $\beta_0$ defined earlier as the average of the propagation constants of S1 and S2 super-modes, $\eta_c$ in the coupling efficiency defined in Eq.~(\ref{eq:coupling_eff}). Tunnelling of generated signal photons to waveguide A is only efficient within a narrow spectral range in a vicinity of the anti-crossing. This is reflected in the overall pre-factor $\eta_c$ of the phase-matching function $\Phi_A$, cf. Fig.~\ref{fig:mode_dispersion}(b). Thus the corresponding photon pair states will be bandwidth limited in the signal photon component, with the bandwidth being entirely controlled by the geometry only. Furthermore, at short propagation distances the two sinc- functions in the expression for $\Phi_A$ will overlap spectrally. This interference can cause a subsequent localization of the phase function $\Phi_A$ in its idler frequency argument. To illustrate this, in Fig.~\ref{fig:filter} we plot $\Phi_A$ as a function of $\Lambda z$ and $\Delta\beta_0 z$ with fixed $\Lambda$ and $\Gamma$ (which corresponds to fixing a signal photon frequency). 

At distances larger than the beat length of the two supermodes, $z>\pi/L$, the phase function has two pronounced peaks at $\Delta\beta_0-\Lambda=0$ (corresponding to phase matching with S1 mode, $\Delta\beta_1=0$) and $\Delta\beta_0+\Lambda=0$ (S2 mode, $\Delta\beta_2=0$). In contrast, when $z<\pi/L$, $\Phi_A$ has a single peak at $\Delta\beta_0=0$. Its largest bandwidth $\Delta\beta_0\approx \pi$ thus implicitly defines the bandwidth in idler photon frequencies.

\begin{figure}
\includegraphics[width = 0.48\textwidth]{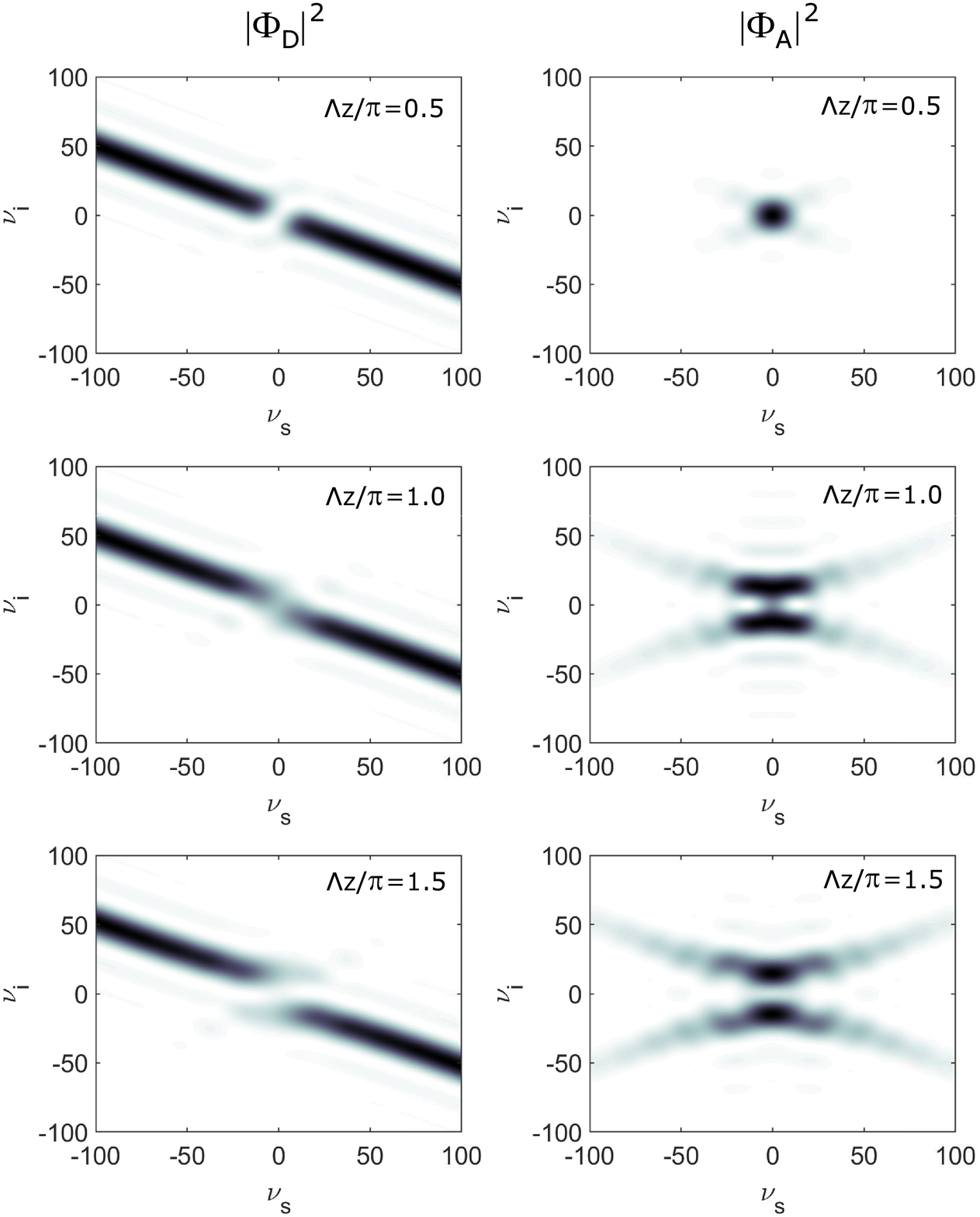}
\caption{Phase functions $\Phi_D$ (left column) and $\Phi_A$ (right column) as functions of dimensionless frequency detunings $\nu=\delta\cdot(\beta_p^\prime z/\pi)$ at propagation distances $z=0.5\pi/\Lambda$ (top row), $z=\pi/\Lambda$ (middle row) and $z=1.5\pi/\Lambda$ (bottom row). Other parameters are the same as in Fig.~\ref{fig:purities}(b).}
\label{fig:PhiA_and_D}
\end{figure}

In Fig.~\ref{fig:PhiA_and_D} $\Phi_D$ and $\Phi_A$ are plotted as functions of dimensionless signal and idler frequency detunings for propagation distances $z=0.5\pi/\Lambda$ (corresponding to one beating length of the signal), $z=\pi/\Lambda$ (two beating lengths) and $z=1.5\pi/\Lambda$ (three beating lengths). $\Phi_D$ reaches its maximum value away from the hybridization point $\nu_s=0$, where it follows the phase matching condition of the isolated D waveguide $\Delta\beta_D=0$. In a vicinity of $\nu_s=0$, where signal photon can couple to the adjacent waveguide, $\Phi_D$ is considerably suppressed and oscillates with $z$. It is easy to see from the expression in Eq.~(\ref{eq:PhiD}), that at the centre ($\nu_s=0,\nu_i=0$), i.e. at the point where $\Delta\beta_0=0$, the phase function oscillates in $z$ as: $|\Phi_D|^2\sim \sin^2(\Lambda z)$.

In contrast, $\Phi_A$ reaches its highest values in the vicinity of the hybridization point $\nu_s=0$, where signal photon coupling to the adjacent waveguide A is most efficient. Consistently with the analysis in Fig.~\ref{fig:filter}, one can observe a qualitative change in the structure of $\Phi_A$ with increasing propagation distance. At distances larger than two beat lengths $\Phi_A$ reaches its maximum along the phase matching curves of the two signal supermodes $\Delta\beta_{1,2}=0$. However at shorter distances, $\Phi_A$ has a single peak at $\nu_s=0, \nu_i=0$, as seen e.g. in Fig.~\ref{fig:PhiA_and_D} for $z=0.5\pi/\Lambda$. Therefore, in this regime, generated photon pairs with the signal photon localized in the waveguide A appear to be spectrally filtered in both signal and idler components.

We note that the phase-matching function $\Phi_A$ is highly factorisable at propagation distances below two beat lengths, $\Lambda z/\pi<1$. In particular, the calculated Schmidt decomposition of $\Phi_A$ at $z=\pi/(2\Lambda)$, top right plot in Fig.~\ref{fig:PhiA_and_D}, gives $P=0.58$. At even shorter distances, this value rises up to $P=0.76$. Therefore with a broadband pump function $\alpha(\omega)$, the factorisability of such photon-pair states is entirely dictated by properties of $\Phi_A$. With no requirement to tune the bandwidth of pump function, one can benefit from using high peak power ultra-short pump pulses to develop high-purity single photon sources with high photon count rates. At the same time, such intrinsic filtering scheme does not suffer from heralding efficiency degrading known for the standard post-filtering scheme even with ideal filters \cite{Meyer-Scott2017}. 

While we used dimensionless frequency detunings $\nu$ in the above plots, the actual bandwidth of the phase-matching function is determined by several geometrical factors: coupling parameter $\Lambda$ (controlled by separation between waveguides D and A), the ratio between interaction length $z$ and beating length (i.e. the overall size of the device), dispersion of group velocity of the pump and the bandwidth of the coupling (both controlled by the selection of modes in waveguides D and A). The multitude of degrees of freedom available for adjustments can be useful for development of practical applications of such setups for intrinsic spectral filtering of photon pairs.

\section{Summary}

Introducing asymmetry in the conventional waveguide coupler setup, we promote a new avenue in design of integrated SPDC-based sources of heralded single photons. With two different arms of the coupler, it is possible to generally suppress coupling except within specific narrow spectral intervals where pairs of modes are engineered to have matching propagation constants. Specifically, we considered the case when a bright light pump and one of the generated photons (idler) remain confined in one arm of the coupler (driven waveguide), while the second photon (signal) can hop into the adjacent arm. We demonstrate that such setup enables several degrees of freedom for advanced manipulation of joint spectral properties of generated photon pairs.

Mode hybridization strongly modifies dispersion of the signal photon, which can be utilized for adjusting the phase matching and shifting the balance between group velocities of interacting waves. The latter is a crucial parameter which affects joint spectral properties of generated photon pairs. As shown in this work, with the help of an auxiliary waveguide, the factorisability of SPDC-produced two photon states can be significantly improved. We note that alternative established techniques to improve purity of heralded single photons rely on either spectral filtering \cite{Meyer-Scott2017}, or apodization of effective interaction along the waveguide length \cite{Branczyk2011}. The advantages of our approach is that it does not affect heralding efficiencies to the same damaging extent as in the case of spectral filtering, and the setup is much simpler and can be much more compact than in typical apodization schemes.

We also demonstrate that interference of two super-modes, in which signal photon is being generated in SPDC process, combined with the pronounced dispersion of coupling, can lead to localization of the joint spectral function of the two-photon states in the 2D space of signal and idler frequencies. In this regime of intrinsic spectral filtering, highly factorisable states can be produced even when using a broadband pump, which can be advantageous for boosting count rates of such heralded single photon sources.

While we focused our discussion on SPDC processes driven by quadratic ($\chi_2$) optical nonlinearities, our ideas can easily be transferred to similar spontaneous four-wave mixing processes in $\chi_3$ structures.

%\bibliography{library}
\bibliography{assymetric_coupler_paper}

\appendix

\section{Hybridized modes in evanescent coupling regime \label{ap:superT}}
The two alternative ways of looking at a waveguide coupler represent two alternative measurement basis with a unitary transformation to convert between them.  In a coupled waveguide basis the operators evolve as coupled harmonic oscillators: 
\begin{equation}
-i\frac{d}{dz}
\begin{bmatrix}
\hat{a}_{D} \\
\hat{a}_{A} \\ 
\end{bmatrix}
= 
\begin{bmatrix}
\beta_{D} & C \\
C & \beta_{A} \\ 
\end{bmatrix}
\cdot
\begin{bmatrix}
\hat{a}_{D} \\
\hat{a}_{A} \\ 
\end{bmatrix}\;.
\label{eq:couple_occ}
\end{equation}
Where $\hat{a}_{D}$ ($\hat{a}_{A}$) creates a signal frequency in the driven (auxiliary) waveguide, $\beta_{D}$ and $\beta_{A}$ are the propagation constants of the two waveguides at infinite separation and C is the coupling constant determined by the evanescent field overlap. 

The transformation T maps these operators into the diagonal super-modes basis:
\begin{equation}
T^\dagger
\begin{bmatrix} 
\beta_{s1} & 0 \\
0 & \beta_{s2} \\ 
\end{bmatrix}
T
= 
\begin{bmatrix}
\beta_{D} & C \\
C & \beta_{A} \\ 
\end{bmatrix}\;,
\label{eq:transf_matrix}
\end{equation}
where $\beta_{s1,s2}$ are eigenvalues of (\ref{eq:couple_occ}):
\begin{equation}
\beta_{s1,s2} = \frac{\beta_D + \beta_A}{2} \pm \sqrt{\frac{(\beta_D - \beta_A)^2}{4} + C^2} \;.
\end{equation}
Apparently, the coupled oscillator model in (\ref{eq:couple_occ}) describes well dispersion of supermodes so long as the splitting of propagation constants remains symmetrical (evanescent coupling regime):
\begin{eqnarray}
\beta_{s1,s2} = \beta_0 \pm \Lambda, 
\label{eq:eigs}
\end{eqnarray}
where $\beta_0 = (\beta_{A} + \beta_{D})/2=(\beta_{s1} + \beta_{s2})/2$, $\Lambda = (\beta_{s1} - \beta_{s2})/2$. Introducing $\Gamma = (\beta_{D} - \beta_{A})/2$, $C^2=\Lambda^2-\Gamma^2$, the normalized eigenvectors can be written as:
\begin{eqnarray}
    \vec{x}_{S1}&=&\frac{1}{\sqrt{2\Lambda}}
    \left[
    \begin{array}{c}
    \sqrt{\Lambda+\Gamma}\\
    \sqrt{\Lambda-\Gamma}
    \end{array}
    \right]\;,\\
    \vec{x}_{S2}&=&\frac{1}{\sqrt{2\Lambda}}
    \left[
    \begin{array}{c}
    \sqrt{\Lambda-\Gamma}\\
    -\sqrt{\Lambda+\Gamma}
    \end{array}
    \right]\;,
\end{eqnarray}
from which the transformation matrix in Eq.~(\ref{eq:T_matrix}) is obtained.

For the input state in the waveguide D $\vec{x}(z=0)=[1,0]^T$, the evolution is given by:
\begin{equation}
    \vec{x}(z)=\frac{e^{i\beta_0 z}}{2\Lambda}\left[
    \begin{array}{c}
     (\Lambda+\Gamma)e^{i\Lambda z}+ (\Lambda-\Gamma)e^{-i\Lambda z}\\
     \sqrt{\Lambda^2-\Gamma^2}e^{i\Lambda z}- \sqrt{\Lambda^2-\Gamma^2}e^{-i\Lambda z}
    \end{array}
    \right]\;,
\end{equation}
such that the amplitude in the waveguide A after one beat length is:
\begin{equation}
    x_A\left(z=\frac{\pi}{2\Lambda}\right)=i\frac{\sqrt{\Lambda^2-\Gamma^2}}{\Lambda}\;.
\end{equation}

Hence the expression for the coupling efficiency in Eq.~(\ref{eq:coupling_eff}) is obtained.

\end{document}